\documentclass[aps,prd,showkeys,showpacs,amssymb,cite,
amsfonts,epsf,preprintnumbers,nofootinbib,superscriptaddress]{revtex4}

\usepackage[dvips]{graphicx}
\usepackage{bm,latexsym,amsmath,amssymb,amsfonts,dsfont}
\usepackage[usenames,dvipsnames]{color}
\usepackage[colorlinks=true,linkcolor=blue]{hyperref}
\usepackage{color}
\usepackage{soul}
\usepackage{epsfig}
\newcommand{\half}{\frac12}
\newcommand{\be}[1]{\begin{equation} \label{#1}}
\newcommand{\ee}{\end{equation}}
\newcommand{\bea}{\begin{eqnarray}}
\newcommand{\eea}{\end{eqnarray}}
\newcommand{\ba}{\begin{array}}
\newcommand{\ea}{\end{array}}
\newcommand{\nn}{\nonumber}

\newcommand{\cL}{\mathcal{L}}

\newcommand{\cP}{\mathcal{P}}

\begin{document}
\title{ 
Entropy of Self-Gravitating Anisotropic Matter
}

\author{Hyeong-Chan Kim}
\email{hyeongchan@gmail.com}

\author{Youngone Lee}
\email{youngone@ut.ac.kr}

\affiliation{School of Liberal Arts and Sciences, Korea National University of Transportation, Chungju 380-702, Korea} 
\begin{abstract}
We examine
 the entropy of self-gravitating anisotropic matter 
confined to a box 
in the context of general relativity.
The configuration of self-gravitating matter 
is spherically symmetric, 
but has anisotropic pressure 
of which angular part
 is different from the radial part.
We deduce the entropy  from the relation 
between the thermodynamical laws and the continuity equation. 
The variational equation 
for this entropy is shown to reproduce
 the gravitational field equation 
for the anisotropic matter.
This result re-assures us 
the correspondence between gravity and thermodynamics.
We apply this method 
to calculate the entropies of a few objects 
such as compact star and wormholes.

\end{abstract}
\maketitle


\section{Introduction}\label{sec:intro}

Ever since Hawking deduced
 the thermodynamics of black holes~\cite{Hawking:1974sw},
thermodynamic approach to 
gravitating systems~\cite{Bekenstein:1974ax,Jacobson:1995ab,Padmanabhan:2009vy,Verlinde:2010hp,Carlip:2014pma}
 provides good insights 
because it considers 
a relatively few thermodynamic variables
 rather than dealing 
complicated dynamic gravitational field equations.
Getting thermodynamic quantities has thus been widely studied
 to understand gravitating systems.
Examining self-gravitating systems has been 
one of those efforts for decades, 
which helps us to understand astrophysical systems. 
Especially, the entropy 
of a spherically symmetric self-gravitating radiation 
and its stability were calculated 
in a series of researches~\cite{cocke,Sorkin:1981}.

Those studies have shown that
requiring maximum entropy 
of self-gravitating radiation in a spherical box
reproduces
 the Tolman-Oppenheimer-Volkoff (TOV) equation for hydrostatic equilibrium~\cite{TOV}.
This equivalence has been dubbed as 
`Maximum Entropy Principle'(MEP).
The total entropy is regarded as 
an action functional of 
mass, $m(r)$, the energy density $\rho(r)$ and the radius $r$.
Varying the total entropy, 
one gets the hydrostatic equation.
Hence, having exact entropy of a system
 enables one to get a hydrostatic equation through MEP.
Recently, there have been discussions 
on the MEP in more general perspectives
\cite{Gao:2011hh,Roupas:2014nea,Cao:2013xy}.
In most of those studies, the gravitating matters 
were assumed to be perfect fluids 
of which pressure $p(r) = w\rho(r)$ is locally isotropic. 
The entropy density for perfect fluids can be deduced 
from the continuity equation 
with series of thermodynamic equation
 combined~\cite{Weinberg:1972kfs}.
It was given by a function 
of the energy density 
and $w$, that is, $s_P\equiv s_P(\rho,w)$. 

When the energy distribution is smooth
in a sufficiently small region of volume $V$,
one can define energy density $\rho$ 
from $U(T,V)=\rho(T)V$
for the total energy $U$ inside the volume. 
Using the first law of thermodynamics, 
$TdS=dU+pdV$,
we have
\bea
dS=\frac{V}{T}\frac{d\rho}{dT}dT+\frac{\rho+p}{T}dV,
\eea
where $T$ and $S$ denote the temperature 
and the entropy of the system respectively.
Since the entropy is a scalar quantity, 
the exactness condition of  $S$ 
determines $\rho$ as a function of $T$:
\bea
\frac{\partial}{\partial T}\frac{\partial S}{\partial V}=\frac{\partial}{\partial V}\frac{\partial S}{\partial T}
~~\longrightarrow~~\rho=\sigma(w) ~T^{\frac{1+w}{w}},
\eea
for a perfect fluid 
with $p=w\rho$, where $w$ is a constant.
The entropy density  of the fluid can be deduced
by considering an isochoric process, for
which $dU = TdS$,
\footnote{
\label{f1}
The first law takes the form $TdS=dU+pdV$ if 
the volume of a system is allowed to change
as in an expanding universe.
Then, the entropy density is given by $s_P=2\alpha_w\rho^\frac{1}{1+w}$.
In either cases,
the entropy density is proportional to 
$\rho^{\frac{1}{1+w}}$ 
up to a constant.
}
\bea
\label{sp}
S=(1+w)\sigma(w)~T^{\frac{1}{w}}V+\mbox{constant}~~\rightarrow~~s_P\equiv\frac{S}{V}=\alpha_w \rho^{\frac{1}{1+w}},~~~\alpha_w=(1+w)\sigma(w)^{\frac{w}{1+w}},
\eea
where the integral constant 
can be set to zero 
by using the third law of thermodynamics.
The constant $\alpha_w$ depends on 
the physical nature of matter consisting the fluid such as $w$.
This entropy density was already given
in Ref.~\cite{Zurek:1984zz,Chavanis:2007kn}.
For radiation with $w=1/3$,
 the density $\rho_{\rm rad}=\sigma_{\rm rad} T^4$ 
and the constant $\sigma_{\rm rad}$ is the Stefan-Boltzmann constant.\footnote{
\label{f2}
$w$ here corresponds to $\gamma-1$ in Ref.~\cite{Zurek:1984zz,Chavanis:2007kn}.
For radiation, 
the Stefan-Boltzmann constant was derived 
by using the Planck's law,
$\sigma_{\rm rad} =\frac{\pi^2 k^4}{60\hbar^3c^2}$.
In Ref.~\cite{Chavanis:2007kn},
 the entropy density was obtained 
for the special cases:
i) the thermal energy is much smaller than the Fermi energy
in the core of neutron stars,
ii) entropy density is proportional to 
number density, $s=\lambda n$, 
where $\lambda$ is a constant.
The entropy density so obtained 
depends on the two constant $\gamma=w+1$ and a constant $K$.
The author calculated $K$ 
for the case of radiation with $w=1/3$.
In these cases,
 the constant $\sigma$ is a function of $\gamma$ and $K$,
which can be calculated 
from the properties of radiation.
It is not clear that
 the values of constant $\sigma$ are the same 
for  all matters with the same $w$.
}

Although majority of researches have focused on 
the dynamics of perfect fluid~\cite{Stephani:2003tm,Delgaty:1998uy,Semiz:2008ny}, 
anisotropic matter in cosmological configuration
 has recently drawn interests~\cite{Ruderman:1972aj,Herrera:1997plx,Bowers:1974tgi,Matese:1980zz,Mak:2001eb,Thirukkanesh:2008xc,Ivanov:2002xf,Varela:2010mf,Bekenstein:1971ej}. 
For example, 
the authors in~\cite{Cho:2017nhx} obtained stable black holes 
with anisotropic matter.
It is well known that, in general relativity, 
the throat of a traversable wormhole needs to be made of 
anisotropic matter~\cite{Morris:1988cz,Cataldo:2016dxq}.
Thus studies on the dynamics of anisotropic matter 
including thermodynamics
 become more important than ever.

For an anisotropic matter 
satisfying the linear equation of state,
\be{eos}
p_k=w_k\rho, \qquad k=1,2,3,
\ee
 the entropy density $s_A$ will depend on 
$\rho,$ and $w_k$ where the subscript $A$ denotes the `anisotropic' nature.
That is, $s_A\equiv s_A(\rho,w_1,w_2,w_3)$. 
When all $w$'s are equal, 
$s_A$ goes to 
that of the perfect fluid.
Since an entropy density is a scalar quantity, 
$s_A$ can be written 
as a product of $s_P$ in Eq.~\eqref{sp} and 
a scalar function $\phi$
that describes the effect of anisotropy:
\bea
\label{sasp}
s_A(\rho,w_1,w_2,w_3)=\phi(w_1-w,w_2-w,w_3-w, \rho)\cdot s_P(\rho,w)
\eea
where $\phi(w_k)\rightarrow 1$ in the isotropic limit.
The function $\phi$ may also contain 
matter information
which is not described by
the equation of state.

In the above discussion,
the pressure $\vec p\equiv (p_1,p_2,p_3)$ is divided
into two parts.
One is the isotropic part of pressure $\vec p_{\rm iso} \equiv p_{\rm iso} (1,1,1)$
and the other is the deviation from it, $\vec{p}_{\rm d} \equiv \vec p-\vec p_{\rm iso}$.
However, there is an arbitrariness 
in the choice of the isotropic part.
The first intuitive choice is 
$ p_{\rm iso}\equiv \bar{p},~\bar{p}=( p_1+p_2+ p_3)/3$.
The other choice could be that
$p_{\rm iso}\equiv p_1$,
and so on.
The resulting entropy should be 
independent of this choice.
This freedom will be
gauged by a compensation vector
later in this work.

We obtain the explicit form of
 the entropy density of self-gravitating anisotropic matter 
confined to a spherically symmetric box of radius $R$ 
from thermodynamic consideration 
up to a constant multiplication factor, 
which depends on 
the individual characteristics of matter.
With this entropy density, 
we perform
 the  variation of total entropy 
of the anisotropic matter 
and show that  
the variational equation reproduces
 a modified TOV equation.
The result of this article shows that 
MEP still holds.
We also seek applications of this entropy 
to cosmological phenomena.

The order of the article is the following:
We review properties of anisotropic matter in Sec.~\ref{sec:amatter}.
In Sec.~\ref{sec:entropy},
 the entropy of an anisotropic matter is obtained 
by comparing the continuity equation 
with thermodynamic relations.
The MEP for a perfect fluid
 is briefly reviewed in Sec.~\ref{sec:maxima}.
The MEP yields the exact dynamical equation (modified TOV) 
using the entropy density of the anisotropic matter obtained in Sec.~\ref{sec:entropy}.
We obtain explicitly the total entropy of a ball of anisotropic matter and apply this calculation to discuss properties of objects such as wormhole and a compact star in Sec.~\ref{sec:app}.
Finally, we summarize our work and discuss about future issues in Sec.~\ref{sec:summary}.

\section{Anisotropic matter}\label{sec:amatter}

The stress-energy tensor $T^{ab}$ of a relativistic matter 
can be  divided into 
a thermostatic part $T_0^{ab}$ 
and a dissipative part $T_1^{ab}$ as
\bea
\label{Tdiv}
T^{ab}=T_0^{ab}+T_1^{ab},
\eea
The dissipative part is written as 
\bea
\label{Tsheer}
T_1^{ab}=u^aq^b+u^bq^a+\pi^{ab},~~~~u_aq^a=0,
\eea
where $u^a$ is
a timelike unit normal vector field.
The quantities $q$ and $\pi^{ab}$ are interpreted as 
a heat flux vector and a viscous shear tensor, respectively.
 The viscous shear tensor satisfies 
$\pi^{ab}u_b=0$ or $u_a\pi^{ab}u_b=0$
 for the Landau frame or the Eckart frame, respectively~\cite{Weinberg:1972kfs}.

The thermostatic part of an isotropic perfect fluid
has the form
\bea T_0^{ab}=\rho u^au^b+p h^{ab}, ~~h^{ab}=x^ax^b+y^ay^b+z^az^b=g^{ab}+u^au^b,
\eea
where $\rho$ and $p$ are the rest frame energy density and  the isotropic pressure, respectively.
The set of four mutually orthogonal vector fields $\{u^a,x^a,y^a,z^a\}$ conforms 
a frame of orthonormal vector fields. 
That is,
\bea
g^{ab}=-u^au^b+x^ax^b+y^ay^b+z^az^b,~~~~
u_au^a=-1,~~x_ax^a=y_ay^a=z_az^a=1.
\eea
When matter is not perfect, 
more general  tensor
 for anisotropic pressure replaces $T_0$,
 for example, 
elastic solid has anisotropic stress tensor \cite{Hayward:1998hb}.
In the absence  of sheer,
off-diagonal components of $\pi^{ab}$ vanishes.
The energy-momentum tensor of the anisotropic matter 
being considered in this article has the form:
\footnote{
The anisotropic matter 
can also be regarded
as an imperfect fluid.
As we mentioned in Sec.~\ref{sec:intro},
there is no 
preferred decomposition of 
the isotropic part in
\eqref{EMtensor}.
For example, 
the isotropic part of 
the pressure
can be $p$ or $p+\pi$:
\bea
[\rho u^au^b+ph^{ab}]+\pi^{ab}
=[\rho u^au^b+(p+\pi)h^{ab}]+\sigma^{ab},~~
\pi\equiv \frac13\mbox{tr}(\pi^{ab}),
\eea
where the terms 
in the bracket represent
the isotropic part and
$\sigma^{ab}$ is 
the traceless part of $\pi^{ab}$,
i.e.,
$\pi^{ab}=\pi(x)h^{ab}+\sigma^{ab}$.
One can find that
$p$ and $\pi$ can have any values
provided that
they satisfy $p+\pi=(p_1+p_2+p_3)/3$.
}
\bea
\label{EMtensor}
T_A^{ab}\equiv
(T_A)_0^{ab}=\rho u^au^b+p_1x^ax^b+p_2y^ay^b+p_3z^az^b,
\eea
where $\rho$ and $p_k(k=1,2,3)$ denote the energy density and
the pressure measured in locally orthogonal rest frame, respectively.

A static, spherically symmetric metric can be written as
\bea
\label{asol}
ds^2=-e^{u(r)}dt^2+e^{v(r)}dr^2+r^2(d\theta^2+\sin^2{\theta}d\varphi^2).
\eea
Because of the spherical symmetry, 
we have $p_\theta=p_\phi$. 
We denote   $p_1\equiv p_r$ and $p_2\equiv p_\theta=p_\varphi$
from now on.
The Einstein equations $G^t_{~t}=8\pi T^t_{~t}$
and $G^r_{~r}=8\pi T^r_{~r}$ read
\bea
\label{Eineq}
e^{-v}&=&1-\frac{2m(r)}{r},~~~~m(r)\equiv \int^r dr'4\pi r^2\rho(r'),\nn\\
u'&=&\frac{2(m+4\pi r^3p_1)}{r^2\left(1-\frac{2m}{r}\right)}.
\eea
With these equations and $G^\theta_{~\theta}=8\pi T^\theta_{~\theta}$
gives the modified TOV equation:
\bea
\label{TOVA}
\frac{dp_1}{dr}=-\frac{\rho+p_1}{r(r-2m)}\left(m+4\pi r^3p_1\right)+\frac{2}{r}(p_2-p_1),
\eea
which resembles the  TOV equation of the isotropic fluid \cite{TOV}
up to the last term:
\bea
\label{TOV}
\frac{dp}{dr}=-\frac{\rho+p}{r(r-2m)}\left(m+4\pi r^3p\right).
\eea

\section{Entropy density of anisotropic matter}\label{sec:entropy}
The temporal part of the continuity equation,
$\nabla_aT_A^{ab}=0$, 
can be interpreted as 
a thermal relation
\bea
-u_b\nabla_aT_A^{ab}=T\nabla_a(s_A u^a)+\mu_N\nabla_a(n u^a),
\eea
here $T, s_A, n$ and $\mu_N$ denote
the temperature, the entropy density,
the number density  and the chemical potential of the particles 
composing the anisotropic matter, respectively.

In the presence of  the dissipative part,
$T_1^{ab}\neq 0$,
the entropy vector $s_A^a$ 
can be defined~\cite{Weinberg:1972kfs,Stephani:1982ac}:
\bea
s_A^a=s_A u^a-\beta u_b(T_{1}^{ab}),
\eea
where  $\beta$
is the inverse temperature 
($\beta\equiv T^{-1}$).
The net entropy production 
can be written as in Ref.~\cite{Schatz:2016gsy},
\bea
\nabla_a s_A^a=-\nabla_{(a}\beta u_{b)}\cdot T_1^{ab}.
\eea
There are two ways for the entropy to be conserved.  
First, there exists a timelike Killing vector field $\xi^a\equiv \beta u^a$.
Second, the dissipation is absent, $T_1=0$ or $T^{ab}=T_A^{ab}$.
In this work, both of the two requirements hold.

When the number of particles does not change,  $\nabla_a(nu^a)=0$, the continuity equation reads:
\bea
\label{cona}
-u_b\nabla_a T_A^{ab}=T\nabla_a(s_Au^a)=0.
\eea
Let us write the entropy density of anisotropic matter as
 $s_A=\phi \cdot s_P$ as in Eq.~\eqref{sasp}.
Here, $s_P\propto \rho^{\frac{1}{1+w}}u^a$
 is the entropy density for a perfect fluid
with an isotropic pressure
$p_{\rm iso}=w\rho$,
and the function
 $\phi\equiv\phi(w_1-w,w_2-w,w_3-w)$ 
describes 
how much the entropy deviates 
from that of the isotropic form.
Hence, $\phi= 1$ when $w_1=w_2=w_3$, 
i.e., when isotropic.

The equation \eqref{cona} becomes
\bea
\label{factor}
\nabla_a s_A^a=0~\rightarrow~\nabla_a(s_P^a)=-\nabla_a(\log\phi) \cdot (s_P^a),
\eea
where $s_A^a\equiv s_Au^a$ and 
$s_P^a\equiv \alpha_w\rho^{\frac{1}{1+w}}u^a$.
In general, 
the energy-momentum tensor \eqref{EMtensor}
can be divided as:
\bea
T_A^{ab}=T_p^{ab}+(p_1-p)x^ax^b+(p_2-p)y^ay^b+(p_3-p)z^az^b,
~~~~T_p^{ab}\equiv(\rho+p)u^au^b+pg^{ab},
\eea
The continuity equation reads
\bea
\label{temp_con}
-u_b\nabla_aT_A^{ab}=0
~\rightarrow~
u\cdot\nabla\rho+(\rho+p)\nabla\cdot u=
\sum_{k=1}^3(p_k-p)(x_k\cdot\nabla x_k)_a
u^a.
\eea
The above equation 
can be written
as a divergence relation of
the entropy density vector 
of a perfect fluid
$s_P\propto\rho^{1/(1+w)}u^a$:
\bea
\nabla_a\left(\rho^{\frac{1}{1+w}}u^a\right)=
\sum_{k=1}^3\frac{w_k-w}{1+w}(x_k\cdot\nabla x_k)_a
\left(\rho^{\frac{1}{1+w_1}}
u^a \right).
\eea
The case when $w=-1 (\rho+p=0)$
will be treated separately.
Comparing this equation
 with  Eq.~\eqref{factor} gives
\bea
\label{dlnp}
\nabla_a(\log\phi)=
\left[
\frac{w_1-w}{1+w}x\cdot\nabla x_a+
\frac{w_2-w}{1+w}y\cdot\nabla y_a+
\frac{w_3-w}{1+w}z\cdot\nabla z_a
\right]+
\Phi_a(x),
\eea
where $\Phi_a(x)$ is the compensation vector mentioned  in Sec.~\ref{sec:intro}.
It  is a vector orthogonal to $u^a$ ($\Phi_au^a=0$)
and represents the anisotropic effect. 
In other words,  $\Phi_a=0$ when matter is isotropic.

For a spherically symmetric configuration,
we can take $x^a,y^a,z^a$ as unit radial, axial and azimuthal spacelike vectors, respectively.
In this case,
 for the metric \eqref{asol},
a calculation in the orthonormal basis 
\bea
\label{basis}
u^a=e^{-u/2}\left(\frac{\partial}{\partial t}\right)^a,
~x^a=e^{-v/2}\left(\frac{\partial}{\partial r}\right)^a,
~y^a=\frac{1}{r}\left(\frac{\partial}{\partial \theta}\right)^a,~z^a=\frac{1}{r\sin\theta}\left(\frac{\partial}{\partial \varphi}\right)^a,
\eea
gives
\bea
\label{xdx}
x\cdot\nabla x_a=0,~~~
y\cdot\nabla y_a=-\frac{1}{r}\delta_{ar},~~~
z\cdot\nabla z_a=
-\frac{1}{r}\delta_{ar}-\cot\theta~\delta_{a\theta}.
\eea
With $w_2=w_\theta=w_\phi$, 
one gets from \eqref{dlnp} and \eqref{xdx},
\bea
\label{lnP}
\partial_t(\log \phi)&=&\Phi_t,\nn\\
\partial_r(\log \phi)&=&
\frac{2}{r}\frac{w_2-w}{1+w}+\Phi_r,\nn\\
\partial_\theta(\log \phi)&=&\cot\theta+\Phi_\theta,\nn\\
\partial_\varphi(\log\phi)&=&\Phi_\varphi.
\eea
Since the factor $\phi$
depends only on the radial coordinate,
one can determine 
the components of $\Phi_a$ to be
$\Phi_t=0,\Phi_\theta=-\cot\theta,\Phi_\varphi=0$ and
$\Phi_r=\Phi_r(r)$
without loss of generality.
Therefore, one function $\Phi_r$ is sufficient to gauge the difference
between two isotropic pressures.
Solving the equations gives
\bea
\label{afactor}
\phi=C(w_2-w)\Phi_w(r)~ r^{\frac{2(w_2-w)}{1+w}} ,
\eea
where $C$ is an integral constant
and $\Phi_w(r)\equiv e^{\int\Phi_r}$ is
 a function to be determined.
We call this function $\Phi_w$
 a compensation factor. 
\
Therefore, in general,
the anisotropic entropy density has the form:
\bea
\label{sa_gen}
s_A=\alpha_w\Phi_{w}(r) \rho^{\frac{1}{1+w}}r^{\frac{2(w_2-w)}{1+w}}.
\eea
The entropy density is 
a function of energy density and 
other related parameters($w$'s here).
Because entropy density is an intensive quantity, 
the $r$ dependence above
seems awkward at first sight.
Actually, 
the entropy density can be written as a function of energy density,
 $s_A\equiv s_A(\rho)$.
One can regard $r$ as 
a function of $\rho$ 
since $\rho$ satisfies
a first order differential equation
\bea
\frac{d\rho}{dr}=f(u(r),v(r),\rho,r)
\eea
for a given metric $u,v$ in Eq.~\eqref{asol}.
Given a specific  $u(r)$ and $v(r)$,
the above equation solves $r$ as 
a function of energy density, $r=r(\rho)$.
\\
If one knows a compensation factor $\Phi_w(r)$
for a choice of isotropic pressure,
$p_{\rm iso}=w\rho$,
the compensation factor $\Phi_{w'}(r)$
for other choice $p'_{\rm iso}=w'\rho$
can be calculated.
By equating entropies
$s_A(w,w_2)=s_A(w',w_2)$,
\bea
\label{swsw1}
\alpha_w\Phi_w(r)\rho^{\frac{1}{1+w}}r^{\frac{2(w_2-w)}{1+w}}
=
\alpha_{w'}\Phi_{w'}(r)\rho^{\frac{1}{1+w'}}r^{\frac{2(w_2-w')}{1+w'}},
\eea
one gets $\Phi_{w'}$ 
in terms of $\Phi_w$.
Obtaining a compensation factor $\Phi_w$
for a specific choice
which gives a modified TOV in Eq.~\eqref{TOVA} is not always 
an easy task.
However,
if there is a choice $w_0$
which makes $\Phi_{w_0}$ be a constant,
a general compensation factor
will have the  form:
\bea
\Phi_w(r)\propto \left[ \rho(r) r^{2(1+w_2)}\right]^{\frac{w-w_0}{(1+w)(1+w_0)}} .
\eea
We will show that,
with a choice 
$\vec p_{\rm iso}=\vec p_1$, i.e. $w_0 = w_1$,
the MEP
with a constant $\Phi$
produces the exact
modified TOV equation.
Consequently, 
one can write down
the entropy density of anisotropic matter 
as:
\bea
\label{sa}
s_A=\alpha_{(w_1,w_2)}~ \rho^{\frac{1}{1+w_1}}r^{\frac{2(w_2-w_1)}{1+w_1}},
\eea
where $\alpha_{(w_1,w_2)}$
is a constant that
does not modify
the equation of motion and
is determined 
by the physical nature of anisotropic matter
as we mentioned above 
in Ref.~\cite{Chavanis:2007kn}.
\\
When $\rho+p_1=0$ or $w_1=-1$,
it is obvious that 
one cannot use the formula \eqref{sa} directly.
Thankfully,
one can use the freedom of choice
for the isotropic part to solve this problem.
As an example,
we obtain an explicit form of
 the compensation factor
for $w_1=-1$ in the next section.

\section{Requiring Maximum Entropy}\label{sec:maxima}

The local maximum of  entropy 
of relativistic matter
coincides with
a dynamically stable equilibrium configuration.
(MEP) \cite{cocke,Sorkin:1981}.
The premises in those papers about
 extrema of total entropy 
 also work here. 
The system being considered in this article is 
static and spherically symmetric.
The only difference is that 
the pressure of matter is anisotropic. 
Thus, all the arguments used in Ref.~\cite{Sorkin:1981} about 
extrinsic curvature $K_{ab}$ of a spacelike hyper-surface $\Sigma$
holds here.

In this section, 
we briefly review the MEP 
for an isotropic matter(perfect fluid) 
and then generalize to the anisotropic matter case.
Finally, for the exceptional case with $w_1=-1$,
we obtain the entropy density 
separately.

\subsection{Isotropic matter case}

The total entropy of the isotropic fluid
 in a spherical box of radius $R$
is given by
\bea
S\equiv \int d\Sigma~ s_P^an_a&=&4\pi\alpha_w\int_0^Rdr~\frac{r^2}{\sqrt{1-\frac{2m(r)}{r}}}\rho^{\frac{1}{1+w}}\nn\\
&=&(4\pi)^{\frac{w}{1+w}}\alpha_w\int_0^Rdr~ m'^{\frac{1}{1+w_1}}r^{2\left(1-\frac{1}{1+w_1}\right)}\left(1-\frac{2m}{r}\right)^{-\half}.
\eea
The total entropy can be regarded as
an action integral
$I=\int_0^R dr~\cL_0[m(r),m'(r);r]$ 
where the corresponding Lagrangian is
\bea
\label{iso_L}
\cL_0[m(r),m'(r);r]\equiv (4\pi)^{\frac{w}{1+w}}\alpha_w m'^{\frac{1}{1+w_1}}r^{2\left(1-\frac{1}{1+w_1}\right)}\left(1-\frac{2m}{r}\right)^{-\half} .
\eea
As expected, 
the Euler-Lagrange equation,
$\frac{d}{dr}\left(\frac{\partial\cL_0}{\partial m'}\right)-\frac{\partial \cL_0}{\partial m}=0,$
gives the original TOV equation \eqref{TOV}.

\subsection{Anisotropic matter when $w_1\neq -1$}\label{maxima_wn1}

For anisotropic matter
with static, spherically symmetric distribution,
in accordance with MEP,
the variation of total entropy 
gives
 the exact modified TOV equation \eqref{TOVA}.
The total entropy of the anisotropic matter 
in a box of radius $R$ is given by
\bea
S_A&=&\int d\Sigma~s_A^an_a\nn\\
&=&4\pi\alpha_{w}\int_0^Rdr \frac{r^2}{\sqrt{1-\frac{2m(r)}{r}}}r^{\frac{2(w_2-w_1)}{1+w_1}}\rho^{\frac{1}{1+w_1}},
\eea
where we use the entropy density in Eq.~\eqref{sa} with $\Phi_w=1$.
In this case,
the action to be extremized 
is given by
\bea
\label{aL}
I&=&\int_0^R dr~\cL(m,m';r),~~~\nn\\
\cL(m,m';r)&\equiv& r^{\frac{2(w_2-w_1)}{1+w_1}}\cL_0(m,m';r)
=(4\pi)^{\frac{w}{1+w}}\alpha_{w}\left(m'\right)^{\frac{1}{1+w_1}} r^{\frac{2w_2}{1+w_1}}\left(1-\frac{2m}{r}\right)^{-\half} , 
\eea
where $\cL_0$ is the Lagrangian
 for the isotropic part in Eq.~\eqref{iso_L}.
The Euler-Lagrange equation
\bea
\frac{d}{dr}\left(\frac{\partial\cL}{\partial m'}\right)-\frac{\partial \cL}{\partial m}=0,
\eea
reproduces the modified TOV equation~\eqref{TOVA} 
where we use the relation $m'=4\pi r^2\rho$.

\subsection{Anisotropic matter when $w_1=-1$}\label{maxima_w-1}

The entropy density 
for anisotropic matter depends on  
the energy density $\rho(r)=4\pi r^2 m'(r)$. 
Thus the Lagrangian for perfect fluid \eqref{iso_L}
 can be generalized to the form:
\bea
\label{trialL}
\cL_{ab}=c(m')^a r^b \left(1-\frac{2m}{r}\right)^{-\half},
\eea
where $a,b,c$ are constants to be determined.
The Euler equation, 
\bea\label{Leq}
\frac{m'^2}{\cL_{ab}}\times \left(\frac{d}{dr}\frac{\partial\cL_{ab}}{\partial m'}-\frac{\partial\cL_{ab}}{\partial m}\right)=0 ~\Rightarrow~
a(a-1)m''+\frac{ab}{r}m'+\frac{(a-1)m'^2-a\frac{mm'}{r}}{r-2m}=0,
\eea
can be rewritten 
as a form of modified TOV equation:
\bea
\label{tov-L}
p_1'=-\frac{\frac{w_1}{1-a}\rho}{r(r-2m)}\left(m+\frac{1-a}{a w_1}4\pi r^3 p_1\right)+\frac{\frac{b}{1-a}-2}{r}w_1\rho.
\eea
Comparing this with
 the modified TOV equation \eqref{TOVA}
one obtains:
\bea
a=\frac{1}{1+w_1},~b=\frac{2w_2}{1+w_1}.
\eea
For $w_1\neq -1$, 
the Lagrangian exactly corresponds to 
the one for the entropy density \eqref{sa}
obtained in Sec.~\ref{sec:entropy},
hence confirms our result.

For $w_1=-1$, the solution was explicitly obtained 
in Ref.~\cite{Cho:2017nhx}. 
It is sufficient that 
we use the results 
from the article:
\bea
\label{Kim_sol}
m(r)=M+\frac{K}{2r^{2w_2-1}},~~~~\rho(r)=-p_1(r)=\frac{(1-2w_2)K}{8\pi r^{2+2w_2}},
\eea
where $M$ and $K$ are constants.
Putting this information into
 Eq.~\eqref{Leq} 
we get $a$ and $b$:
\footnote{
$m(r)-M$ instead of $m(r)$ 
satisfies Eq~\eqref{Leq}.
If one solves Eq~\eqref{Leq} directly
one gets $m(r)\propto r^{1-2w_2}$,
after that 
one can freely add
a constant mass.
 This means $m(r)$ in Eqs.~\eqref{trialL} and \eqref{Leq} is related to the physical
mass up to an additional constant.
}
\bea
a=1-\frac{1}{2w_2}, ~~~b=2(a-1)w_2.
\eea
Therefore, the entropy density for $w_1=-1$ which satisfy 
the maximum entropy principle is
\bea
\label{sw-1}
s=s_0 ~\rho^{1-\frac{1}{2w_2}}r^{-\frac{1+w_2}{w_2}},
\eea
where $s_0$ is a constant.

As we discussed 
in the previous section, 
one can determine 
`the compensation factor' $\Phi_w(r)$
for the above entropy density.
When $w_1=-1$,
to avoid the denominator to vanish,
one can choose
the isotropic part
of the pressure as
$p_{\rm iso}\equiv p_1-2p_2$ or $w_{\rm iso}=w_1-2w_2(=-1-2w_2)$.
The entropy density 
using the anisotropic factor \eqref{afactor}
is obtained to be 
\bea
s~\propto ~\Phi_{w}(r)\rho^{-\frac{1}{2w_2}}r^{-\frac{(1+3w_2)}{w_2}},
\eea
by putting $w=-1-2w_2$ to Eq.~\eqref{sa_gen}.
After requiring MEP, 
one obtains the factor $\Phi_w(r)=r^{-2w_2}$
 up to a multiplicative constant.
By using the relation
$\rho(r)\propto r^{-2(1+w_2)}$
in Eq.~\eqref{Kim_sol},
one reproduces
the entropy
in Eq.~\eqref{sw-1}.

\section{Total Entropy and Applications}\label{sec:app}

Total entropy of self-gravitating radiation 
confined to a box of radius $R$
was obtained in Ref.~\cite{Sorkin:1981}.
By using the same method, 
we  obtain a general form of 
total entropy of anisotropic matter
confined to the box:
\bea
\label{totalS}
    S=\left.S_A(r)\right|^{r=R}_{r=0}
~=~\left.\frac{(4\pi)^{\frac{w_1}{1+w_1}}\alpha_{w}}{(1+w_1+2w_2)}\frac{\frac{m}{r}+w_1\frac{dm}{dr}}{\left(\frac{dm}{dr}\right)^\frac{w_1}{1+w_1}\left(1-\frac{2m}{r}\right)^\half}r^{\frac{1+w_1+2w_2}{1+w_1}}\right|^{r=R}_{r=0}.
\eea 
One can verify 
that the derivative of the function 
on the right-hand side
is the Lagrangian \eqref{aL} 
by using the relation~\eqref{Leq}.
We call the indefinite integral $S_A(r)$ 
as `entropy function'.
One can see that 
the above total entropy reproduces
 that in Ref.~\cite{Sorkin:1981}
when $w_1=w_2=w_3=1/3$.

In a general case, solutions of the Einstein equation 
can be singular at the origin.
In order for a solution to be regular
at $r=0$, $m(r)\sim (4\pi r^3/3)\rho(0)\sim r^3$,
and $\rho(r)\sim r^\alpha$ with $\alpha> 0$.
Otherwise the solution will be singular at the center.
To analyze the singularity more explicitly, 
one needs to have an exact solution.
The properties of solutions 
with $w_1=-1$ at $r=0$
 is extensively analyzed
 in Ref.~\cite{Cho:2017nhx} 
with the exact solutions in Eq.~\eqref{Kim_sol}.

The entropy function above 
can be further simplified to
\bea
\label{entropy_r}
S_A(r)
&=&\frac{m+4\pi r^3p_1}{\left(1-\frac{2m}{r}\right)^\half}\left(\frac{s_A(r)}{\rho+p_1+2p_2}\right)
\eea
by using $m'=4\pi r^2\rho$ 
and the relation~\eqref{eos} and \eqref{sa}.

Recently,
in Ref.~\cite{Kim:2017hem} 
the author suggested conditions to avoid  naked singularities
for a system with an anisotropic fluid.
One of the conditions is
 $\rho+p_1+2p_2<0$,
which indicates
 a violation of  the strong energy condition.
Putting this into
the relations \eqref{totalS} or \eqref{entropy_r},
the entropy function takes a negative value.
Since entropy is defined 
by a possible number of physical configurations of a system,
negative value of entropy seems unreasonable.
If we look into 
the derivation of total entropy 
carefully,
the total entropy,
$\int_a^b\cL_A=S_A(b)-S_A(a)$,
is always non-negative
since the entropy function $S_A(r)$
is monotonically increasing function of $r$
even if $S_A(r)<0$.
It is because 
the Lagrangian density, $\cL_A=S'_A(r)$,
is positive.
Therefore,
even though
$\rho+p_1+2p_2<0$,
the total entropy of the system 
is positive.

A wormhole is a solution of general relativity
 that has a throat 
and two sides of entrance and exit~\cite{Weyl:1921,Misner:1957mt,Kim:2001ri}.
In general relativity, 
the throat of a traversable wormhole 
needs to be made of 
exotic, anisotropic material
which violates energy conditions
(especially  null energy condition~\cite{Raychaudhuri:1953yv} 
along with weak, strong and dominant energy conditions).
If anisotropic matter
(that creates  a wormhole)
 is bound ($r<R$)
and the wormhole's throat is located at $r=B$, 
the entropy of the wormhole 
will be given by
\bea
S =2[S_A(R)-S_A(B)],
\eea
provided that
both sides of the wormhole  
have the same shape.

Morris and Thorne \cite{Morris:1988cz} suggested 
specific wormhole solutions 
which use exotic material minimally:
i) a zero-tidal force solutions, 
ii) a solution with a finite radial cutoff of the stress-energy 
and iii) a solution with exotic matter 
limited to the throat vicinity. 
The first and the second solutions satisfy
 the equation of state 
of type $p_k=w_k\rho$ with 
$1+w_1+2w_2=0$.
Although the equation of state
 of the third solution
does not have the linear form,
the matter still satisfy 
$\rho+p_1+2p_2=0$.
Putting these conditions into 
the total entropy formula \eqref{totalS},\eqref{entropy_r},
 one immediately gets
 indefinite  value 
for the entropy function.
From the Einstein equation
 for the metric \eqref{asol}, one gets
\bea
8\pi(\rho+p_1+2p_2)=\left(1-\frac{2m}{r}\right)\left(\frac{2}{r}u'-\frac{v'u'}{2}+\frac{u'^2}{2}+u''\right).
\eea
When $u'=0$ throughout the region 
where exotic matter is used,
 $\rho+p_1+2p_2=0$.
The three examples 
in Ref.~\cite{Morris:1988cz}
 is the case.
For the same metric,
the entropy function can be written as
\bea
\label{Sru}
S_A(r)=
\frac{4\pi r^2 }{\left(1-\frac{2m}{r}\right)^\half}\frac{s_A}{\left(\frac2r-\frac{v'}{2}+\frac{u'}{2}+\frac{u''}{u'}\right)},
\eea
where we use the relation \eqref{Eineq}.
We may slightly  perturb
 the above wormhole solutions
 to avoid the divergence of the entropy function.
If we compare two configurations of
$S(u'=0)$ and $S(|u'|\neq 0)$,
one may tell
which configuration is
more favourable than the other.
For a smooth function 
 $u'(r)~(|u'|\ll1$ with dimension of inverse length)
near the throat vicinity,
the quantities
$(u')^2,u''$
are much smaller than $u'$.
The denominator of the above equation reduces to
\bea
\mbox{The denominator}\propto~~
e^{v}\frac{2r-3m-rm'}{r^2}
+\frac{u'}{2}.
\eea
The sign of the first term
may switch from positive to negative and vice versa.
For example,
 in the third wormhole in Ref.~\cite{Morris:1988cz}
the mass was given by  
$m(r)=b[1-(r-b)/a]^2$,
where $u'=0$
in the region $b\le r\le b+a$.
Provided that
perturbations around $u'= 0$ is small and
the solution of $m(r)$
has the same  quadratic form,
occasional sign changes appear
for various values of $a$ and $b$.
The entropy function
becomes extremely large
when the denominator approaches zero.
When there is a sign change,
the indefinite value
makes comparison difficult.

As we discussed in 
Sec.~\ref{sec:entropy},
the freedom of choice
in isotropic pressure part
may help us to prevent from this pathology.
When we set
the isotropic pressure part as 
$\cP\equiv p_1+2p_2$,
the process of getting the anisotropic factor $\phi$
suffers the same pathology when 
$\rho+p_1+2p_2=0$ or
$w_1+2w_2=-1$
as in the case when $w_1=-1$.
We expect that
with proper `compensation factor' $\Phi_w$
one can obtain 
finite form of entropy density
by using a similar process 
in Sec.~\ref{maxima_w-1}.

The thermodynamic stability of the configuration $u'=0$ can be 
estimated by
the difference of the entropy functions,
$\Delta S=[S_A(r,u'=0)-S_A(r,u'\neq 0)]^b_a$.
To this end,
further analysis with
wider range of  exact solutions
including $u'\neq 0$ will be necessary.

Incidentally, 
we get another form of
the entropy function 
\bea
S_A(r)=
-4\pi\frac{(m+4\pi r^3p_1)}{\sqrt{1-\frac{2m}{r}}~R^0_{~0}}s_A.
\eea
using
$R^0_{~0}=-4\pi( \rho+p_1+2p_2)$.

\section{Summary and Discussions}\label{sec:summary}

In this article
we considered a system of 
self-gravitating, static and spherically symmetric matter
of which pressure satisfies
the linear equation of state 
$p_1=w_1\rho$ and $p_2=p_3=w_2\rho$.
We obtained a form of entropy density 
for the anisotropic matter.
To obtain the exact form of
entropy density,
we introduced a scalar function $\phi(x)$, 
the anisotropic factor
that contains the information
of anisotropic deviation from an isotropic perfect fluid.
The choice of an isotropic part of the pressure was shown to have
an arbitrariness.
A compensation vector was obtained
to gauge this arbitrariness.
This gauge freedom could be beneficial
when we meet a problem.
In the process,
the Maximum entropy principle(MEP) played a key role
in obtaining the exact form of the entropy.
We showed that 
the requirement of local maximum of total entropy gives 
the exact Einstein equation of the system
composed of the anisotropic matter.
This observation supports
the correspondence between thermodynamics and gravity.
By the guidance of this correspondence,
the entropy density
for  $p_1=-\rho$ or $w_1=-1$ 
was obtained separately.

The entropy density  in this article has 
a multiplicative factor that describes 
the anisotropic effect.
The quantity we obtained is 
a function of radius $r$ and $w_2-w_1$.
This comes from the use of the spherical coordinate system 
and the simple form of anisotropy 
between the radial and the angular directions.
In a more general situation, 
the entropy density is expected 
to have various forms as a function
of energy density and pressure.

With the explicit form of the total entropy function \eqref{totalS},
as an application,
we tried to analyze a wormhole solution
by estimating the entropy in the vicinity of a throat.
Actually, most of the issues about wormholes
come from the fact that
matter composing wormholes 
are not ordinary but exotic.
There are various ways to overcome or
 to go around this issue.
For a modified gravity theory, 
one may find wormholes 
without exotic matter.
There were many articles in which
a throat of traversable wormholes 
is composed of 
ordinary matter
in various modified gravity theories,
e.g., Einstein-Gauss-Bonet theory \cite{Mehdizadeh:2015jra}
 or Lovelock theory \cite{Zangeneh:2015jda}, 
$f(R)$ gravity \cite{Mazharimousavi:2016npo},
etc.\cite{Nandi:1997mx,Eiroa:2008hv}.
Similar analysis on the entropy
for the wormholes in those theories
is worth trying.
Especially, 
the method in this article
can be applicable
to an anisotropic matter in the context of $f(R)$ gravity,
because the MEP was proven for self-gravitating fluid
in $f(R)$ gravity~\cite{Fang:2015pcw}.
On the other hand,
 if we stick to the Einstein gravity,
our result shows that 
building materials or conditions
 should be chosen carefully.

There are many studies on astro-physical objects
such as relativistic stars, neutron stars, etc.
in which anisotropic pressure takes important roles
\cite{Herrera:1997plx,Mak:2001eb,Isayev:2017rci}.
We expect that 
the entropy obtained in this article
can be used to estimate
the state of a relativistic star
by comparing the values of entropy 
for different configurations of the star.
The self-gravitating solutions
for the anisotropic matter
were classified in Ref.~\cite{Kim:2016jfh}.
For the case of matter with non-linear
equation of states,
the analysis will be quite difficult and
deeper studies on the entropy formula
 will be required.

\section*{Acknowledgment}
This work was supported by the National Research Foundation of Korea grants funded by the Korea government NRF-2017R1A2B4008513.
Y. Lee is grateful to Dr. Ma, Chung-Hyeun for his kind treatment in the hospital.

\end{document}